\documentclass[english,aps,prb,superscriptaddress,twocolumn]{revtex4-2}
\usepackage{babel}
\usepackage{amsmath,mathtools}
\usepackage{amssymb}
\usepackage{graphicx}
\usepackage{xcolor}

\begin{document}
\title{ Unconventional node voltage accumulation in generalized topolectrical circuits with multiple asymmetric couplings }
\author{S M Rafi-Ul-Islam }
\email{e0021595@u.nus.edu}
\selectlanguage{english}%
\affiliation{Department of Electrical and Computer Engineering, National University of Singapore, Singapore}
\author{Zhuo Bin Siu}
\email{elesiuz@nus.edu.sg}
\selectlanguage{english}%
\affiliation{Department of Electrical and Computer Engineering, National University of Singapore, Singapore}
\author{Haydar Sahin}
\email{sahinhaydar@u.nus.edu}
\selectlanguage{english}%
\affiliation{Department of Electrical and Computer Engineering, National University of Singapore, Singapore}
\affiliation{
Institute of High Performance Computing, A*STAR, 138632, Singapore}
\author{Ching Hua Lee} 
\email{phylch@nus.edu.sg}
\selectlanguage{english}%
\affiliation{Department of Physics, National University of Singapore, Singapore}
\author{Mansoor B. A. Jalil}
\email{elembaj@nus.edu.sg}
\selectlanguage{english}%
\affiliation{Department of Electrical and Computer Engineering, National University of Singapore, Singapore}
\begin{abstract}
A non-Hermitian system is characterized by the violation of energy conservation. As a result of unbalanced gain or loss in the forward and backward directions  due to non-reciprocal couplings, the eigenmodes of such systems exhibit extreme localization, also known as non-Hermitian skin effect (NHSE). This work explores unconventional scenarios where the interplay of multiple asymmetric couplings can cause the NHSE to vanish, with the admittance spectra taking identical dispersion under open boundary conditions (OBC) and periodic boundary conditions (PBC). This is unlike known non-Hermitian models where the NHSE vanishes only when the non-Hermiticity is turned off. 
We derive general conditions for the NHSE, with the overall eigenmode localization determined by the geometric mean of the cumulative contributions of all asymmetric coupling segments. In the limit of large unit cells, our results provide a route towards the NHSE caused by asymmetric hopping textures, rather than single asymmetric hoppings alone. Furthermore, our generalized model can be transformed into a square-root lattice simply by tuning the coupling capacitors, where the topological edge states occur at a non-zero admittance, in contrast to the zero-admittance states of conventional topological insulators.  We provide explicit electrical circuit setups for realizing our observations, which also extend to other established platforms such as photonics, mechanics, optics and quantum circuits.   
\end{abstract}
\maketitle
\section*{Introduction}
Nonhermicity has brought about a plethora of interesting new phenomena \cite{li2021non,lv2021curving,alvarez2018topological,lee2020exceptional,torres2019perspective,yao2018edge,zhao2019non,gong2018topological,geier2021non,li2020quantized,hatano1996localization,leykam2017edge,bergholtz2021exceptional,shen2018topological}, of which the non-Hermitian skin-effect (NHSE) \cite{song2019non,xiao2020non,bhargava2021non,lee2020unraveling,lee2019anatomy,luo2021unifying,roccati2021non,zhang2020correspondence,li2020critical,rafi2021topological,kawabata2020higher,li2020topological} (i.e, extreme localization of the eigenstates to a boundary) has galvanized various reformulations of the conventional concepts of the Brillouin zone (BZ) and bulk-boundary correspondence (BBC) \cite{helbig2020generalized,kunst2018biorthogonal}. While the BBC can already be broken with a single asymmetric non-Hermitian coupling, the more interesting interplay between multiple asymmetric non-Hermitian couplings has not been thoroughly explored. The interplay of multiple dissimilar and possibly asymmetric couplings becomes especially physically relevant in the topolectrical (TE) circuit \cite{hofmann2020reciprocal,albisetti2018nanoscale,ni2021higher,stegmaier2021topological, zou2021observation,song2020realization,zhang2020topolectrical,hofmann2020reciprocal,lee2020imaging,rafi2020realization,rafi2020anti}  context, where circuit connections can be engineered and reconfigured in arbitrarily complicated manners. A variety of novel phenomena associated with non-Hermitian systems has generated much recent interest for novel applications such as ultra-sensitive sensors \cite{hokmabadi2019non,budich2020non}, quantum computations \cite{zhao2020anomoulas,yoshida2019non,pan2018photonic}, quantum Hall states \cite{yao2018non,sun2020spin,chen2018hall,yoshida2019non,sun2019field}, and reflectionless transmitters \cite{zhang2013momentum,wu2018unidirectional}. Hence, the theoretical investigation and experimental implementation of non-Hermitian systems have been extended to various synthetic platforms such as metamaterial \cite{schomerus2020nonreciprocal,zhou2020non}, photonics \cite{silveirinha2019topological,feng2017non}, mechanical \cite{yoshida2019exceptional}, optical \cite{zhang2018non,el2019dawn}, superconducting \cite{zhou2020non,wang2021majorana}, and acoustic \cite{gao2021non,zhu2018simultaneous} systems. 
However, because this is a budding field, most non-Hermitian phenomena have only been experimentally probed through the simplest models and due to the  limited ability to dynamically modulate system parameters,  the study of non-Hermitian phenomena  in the abovementioned synthetic systems are confined mostly to models with single asymmetric couplings and intrinsic loss and gain terms. As a result, significant questions pertaining to the NHSE remain unanswered such as the role of multiple asymmetric couplings on the non-Hermitian skin mode distributions. In particular, it is predicted that the interplay between multiple but dissimilar directional coupling branches may modify the eigenstates localization. Meanwhile, the unprecedented flexibility in tuning the model parameters and types of couplings in an electrical circuit array makes TE the most suitable platform to study the many  unconventional non-Hermitian topological phases in a system with multiple asymmetric branches.  Because circuit platforms possess unparalleled versatility for the simulation of a large varieties of novel phases, it is timely and important to systematically study the generalized NHSE models that can be realized in such systems.

In this paper, we theoretically characterize the various non-Hermitian skin effects in a generalized circuit lattice with multiple asymmetric couplings that break the Hermiticity of the systems. Specifically, we uncover a little-known phase boundary in the non-Hermitian parameters characterized by the complete cancellation of the skin modes such that the admittance  spectra (analogous to the energy dispersion in condensed matter \cite{rafi2020topoelectrical,rafi2020strain,rafi2020anti}) under open boundary conditions (OBC) and periodic boundary conditions (PBC) have identical dispersions. Any deviation of the non-Hermitian parameter from its critical value will cause the NHSE to re-emerge where the boundaries the eigenmodes are localized at determined by whether the non-Hermitian parameter is greater or smaller than the critical value. Furthermore, we deduce a general condition for the extensive localization of overall eigenstates based on the geometric mean of the left- and right- going  couplings of all asymmetric branches, which is distinct from the established criteria for  non-Hermitian systems with single asymmetric branch. Furthermore, we can transform our circuit to a square-root topological lattice system in which we have topological edge modes at non-zero admittance in addition to the bulk skin modes. Our unconventional mechanism of accumulation and cancellation of NHSE via tuning non-Hermitian parameters promises a new pathway to many other exotic topological phenomena. 
\section*{Results}
\subsection{Generalized conditions for NHSE with multiple asymmetric couplings}
The presence of unbalanced directional couplings between the electrical nodes in a topolectrical (TE) circuit lattice leads to different admittance spectra under periodic boundary condition (PBC) and open boundary condition (OBC).  In general, such asymmetric couplings result in the exponential localization of the eigenmode voltage distribution near one of the two terminal nodes of an open chain, as a consequence of the non-Hermitian skin effect (NHSE). This exponential localization occurs because of the shift of the wavevector from a real value, $k \in \mathbb{R}$ under PBC to a complex value $k+i \kappa$ under OBC necessitated by satisfying the boundary conditions at both boundaries. Here, $\kappa$ is associated with the non-Bloch factor given by $\beta = \alpha e^{i k}= e^{i k} e^{-\kappa}$ such that the non-Bloch multiplication factor $\alpha = e^{-\kappa}$ denotes exponential decay of the voltage and $\alpha$ is the non-Bloch multiplication factor. A value of  $\kappa\neq 0$ will invalidate the usual bulk-boundary correspondence and modify the definition of topological invariants, and necessitate the characterization of the properties of the OBC system using the generalized Brillouin zone (GBZ). The GBZ of a non-Hermitian  circuit Laplacian ($L (\beta)$) can be obtained from the solutions of the characteristic equation, which can be expressed as
\begin{equation}
\mathrm{det} [L (\beta)- \mathcal{J} ] =0
\label{eqa}
\end{equation}  
where $L(\beta)$ is the Laplacian of the PBC TE circuit written in Bloch form with $\exp(ik) \rightarrow \beta$, and $\mathcal{J}$ is the admittance eigenvalue. Denoting the size of the matrix $L(\beta)$ as $N$, and arranging the solutions of $\beta$ for a given $\mathcal{J}$ in ascending orders of magnitude such that $|\beta_1| \leq |\beta_2| \leq \ldots \leq |\beta_N|$, the GBZ is given by the loci of $\beta$ where the two $\beta_i$s closest to unity have the same magnitude for some $\mathcal{J}$. 
The NHSE has been previously investigated in the non-Hermitian Nelson-Hatano model $L_{\mathrm{isolated}}(\mathrm{\beta}) = C^{\mathrm{right}}\beta + C^{\mathrm{left}}/\beta$, where $C^{\mathrm{left}}$ and $C^{\mathrm{right}}$  denote the left and right asymmetric nonreciprocal coupling in the unit cell between neighboring nodes \cite{longhi2019topological}. For this system, the decay rate of the OBC chain can be characterized by $\kappa_{\mathrm{isolated}}= \log \sqrt{(C^{\mathrm{right}}/C^{\mathrm{left}})}$.  The presence of only a single node in the unit cell of the model motivates us to investigate how the localization of the voltage eigenstate will be affected when the unit cell contains multiple non-reciprocal couplings. 
To answer this question, we generalize the original Nelson-Hatano model to include $N$ nodes per unit cell with non-reciprocal couplings between neighboring nodes. The Laplacian of the system is then given by 
\begin{align}
L\left(e^{ik}\right) =&\left( \sum_{i}^{N-1} (C^{\mathrm{right}}_i |i\rangle  \langle i+1| + C^{\mathrm{left}}_i|i+1\rangle  \langle i|) \right)\notag\\
	&+  C^{\mathrm{right}}_Ne^{ik}|N\rangle   \langle 1| + C^{\mathrm{left}}_Ne^{-ik}|1\rangle  \langle N|. 
	\label{LbetaC} 
\end{align}
where $|i\rangle$ corresponds to the $i$th node, and $C^{\mathrm{left}}_i$ and $C^{\mathrm{right}}_i$, which are real, denote the couplings of the $i$-th node to its left and right neighbors respectively. (We assume that the circuit has been appropriately grounded so that the diagonal terms are zero, as described in detail later.) 
Eq. \eqref{LbetaC} can be interpreted as a series of $n$ Nelson-Hatano segments with different values of $C^{\mathrm{left}}_i$ and 
$C^{\mathrm{right}}_i$ chained together in series in a periodic manner.
We then find (see Methods) that the overall non-Bloch factor is given by 
\begin{equation}
\alpha_{\mathrm{total}}=\sqrt{\prod_{i=1}^n \frac{|C^{\mathrm{right}}_{i}|}{|C^{\mathrm{left}}_{i}|}}. \label{eqb}
\end{equation}   
Therefore, the NHSE localization of the eigenmodes does not depend on a single asymmetric coupling rather on the resultant contributions from the non-reciprocal couplings between all the pairs of neighboring nodes. More explicitly, the overall inverse decay length or the skin decay rate is the product of the non-reciprocal couplings between those of individual pairs of neighboring nodes in the unit cell.  As a result, there is a competition between which edge the eigenmodes would be localized at in the OBC circuit when $C^{\mathrm{left}}_i/C^{\mathrm{right}}_i > 1$ for some values of $i$, and $C^{\mathrm{left}}_i/C^{\mathrm{right}}_i < 1$ for others.  The localization of the voltage eigenmodes is determined by the value of $\alpha_{\mathrm{total}}$. The eigenmodes are localized to the left (right) edge when $\alpha_{\mathrm{total}}$ is less than (greater than) 1.
More interestingly, the competition between the various asymmetric segments can give rise to a critical condition in which the usual bulk boundary correspondence (BBC) is restored even when the non-reciprocities are present. Let us introduce $C_i \equiv (C^{\mathrm{left}}_i + C^{\mathrm{right}}_i) / 2 $ and $C_{\mathrm{n}i} \equiv  (C^{\mathrm{right}}_i - C^{\mathrm{left}}_i) / 2$. The $C_i$'s are then the Hermitian parts 
of the inter-node coupling terms in Eq. \eqref{LbetaC}, while the $C_{\mathrm{n}i}$s are non-Hermitian parameters representing the deviations of the coupling from reciprocity.  The critical value of one of the non-Hermitian parameters, say $C_{\mathrm{n}1}$, at which the BBC is restored can be expressed as a function of other parameters: $C^{\mathrm{critical}}_{\mathrm{n}1}=f(C_1, C_2,\ldots, C_n, C_{\mathrm{n}1}, \ldots, C_{\mathrm{n}N})$ (see section B for an explicit expression).  At the critical value, the  NHSE vanishes, and the OBC and PBC admittance spectra overlap. The existence of a non-zero critical value for the coupling is in stark contrast to the non-Hermitian behavior of an individual non-reciprocal segment, in which the BBC is respected only when the non-Hermitian parameter is zero.

\subsection{Skin mode accumulation of multiple non-reciprocal segments}
\begin{figure*}[ht!]
\centering
\includegraphics[width=0.9\textwidth]{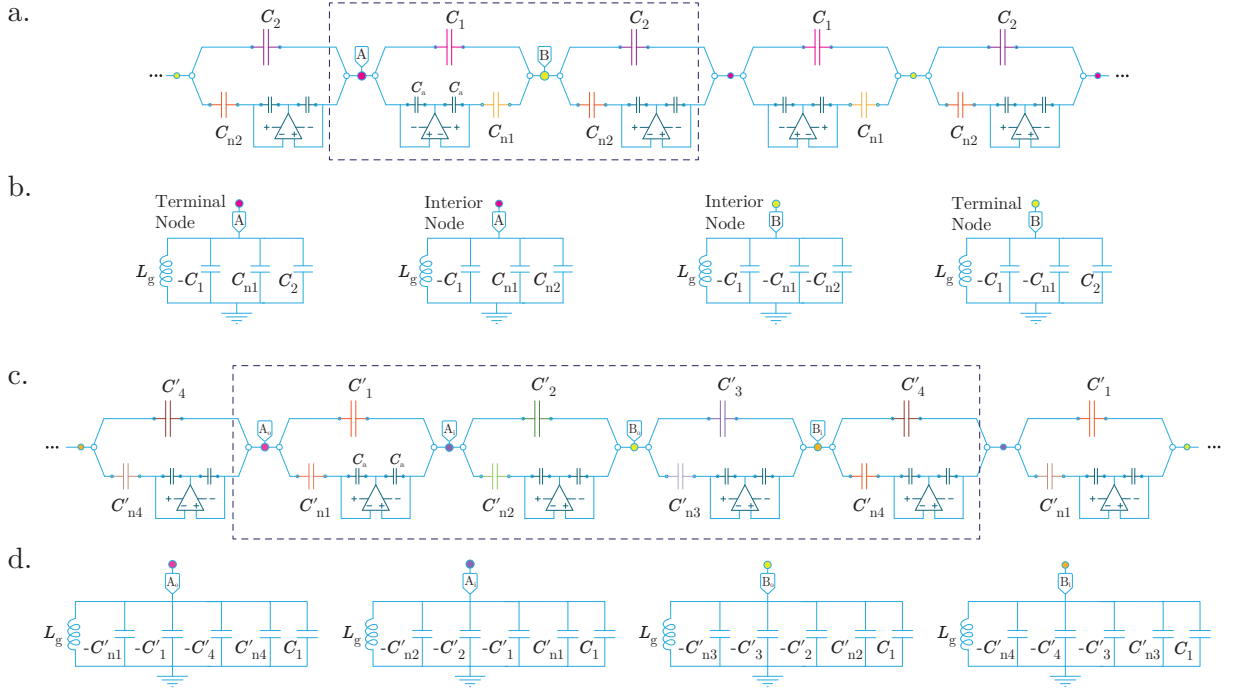}
\caption{ Schematic of the general circuit described by Eq.~\eqref{LbetaC}  with multiple voltage nodes coupled asymmetrically to their neighbors per unit cell. a. A non-reciprocal circuit chain with two nodes labelled $A$ and $B$ per unit cell. The unit cell is marked in a dotted rectangle. The intra- and inter-cell couplings are  $(C_1-C_{\mathrm{n}1})$ and $(C_2+C_{\mathrm{n}2})$ in the forward direction and $(C_1+C_{\mathrm{n}1})$ and $(C_2-C_{\mathrm{n}2})$ in the backward direction, respectively. The coupling imbalance is realized using INICs implemented using OP-amps (LTspice model no LT1056). b. The grounding of the $A$ and $B$ nodes with grounding capacitors and inductors. Note that under OBC, the grounding of the two terminal nodes requires a different configuration from the nodes in the interior of the chain to maintain the same diagonal Laplacian matrix elements throughout the Laplacian matrix. c. The circuit lattice with four asymmetric segments per unit cell. Note that here, we have now set a uniform biasing to the op-amps across all four nodes to simplify the resulting equations. This circuit can be represent a square-root TI when the coupling capacitances are related by certain conditions (see section C for details). d. The grounding connections of the circuit nodes of the chain shown in c.  }
\label{gFig1}
\end{figure*}
\begin{figure*}[ht!]
\centering
\includegraphics[width=0.6\textwidth]{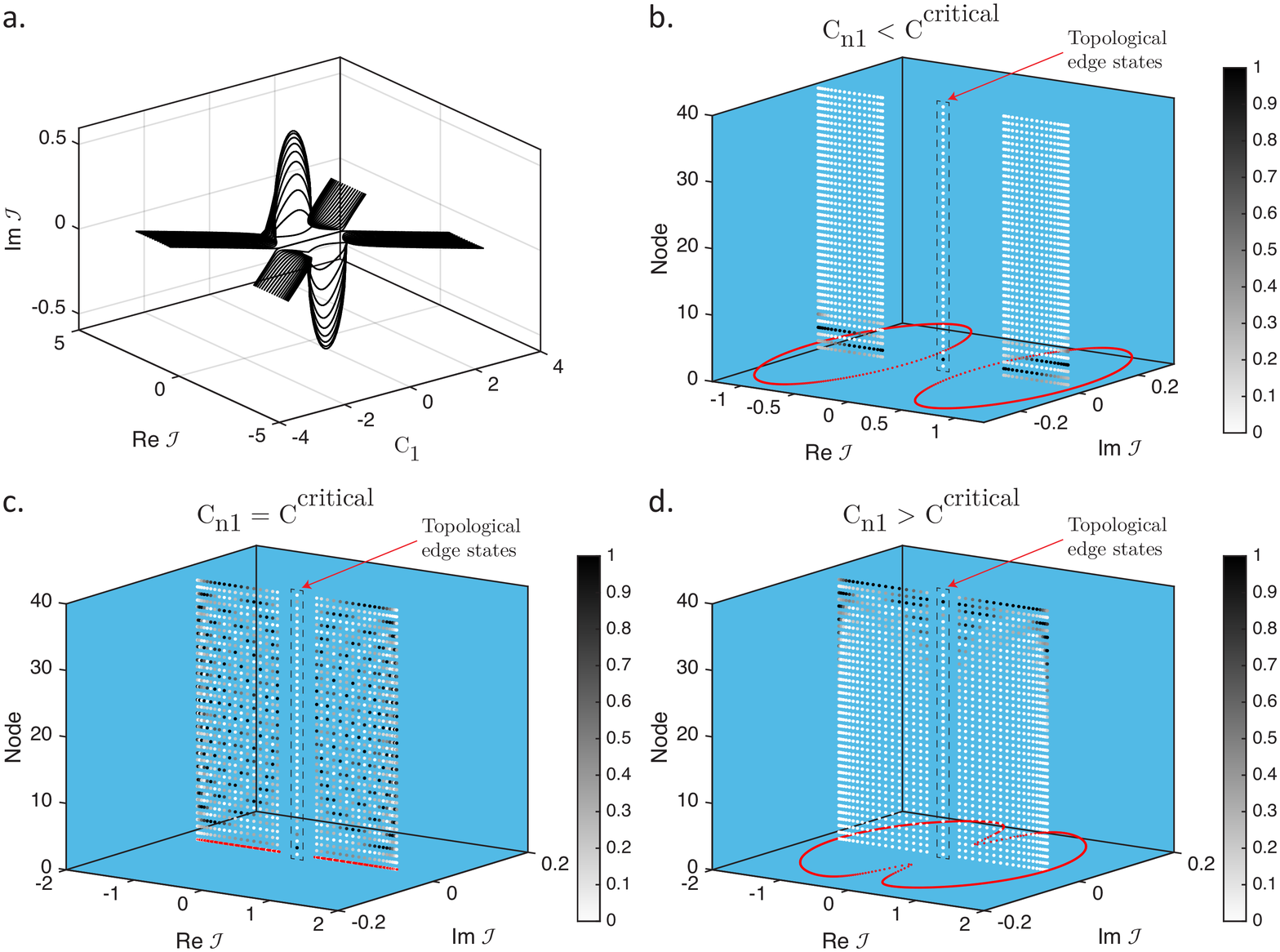}
\caption{Admittance dispersion and NHSE evolution of the circuit arrays with two segments (generalized SSH circuit) described by Eq. \ref{eq1} at resonant frequency. a. OBC admittance spectra at $C_2=1$ $\mu\mathrm{F}$,  $C_{\mathrm{n}1}=0.6$ $\mu\mathrm{F}$ and $C_{\mathrm{n}2}=0.5$ $\mu\mathrm{F}$. b. The OBC and PBC admittance spectrum when $C_{\mathrm{n}1}$ is set less than $C_{\mathrm{n}1}^{\mathrm{critical}}=C_2^{-1} C_1 C_{\mathrm{n}2}$. The red dots on the $\mathrm{Re}\mathcal{J}$ and $\mathrm{Im}\mathcal{J}$ plots indicate the OBC admittance spectra. Each column of white to gray dots on  indicates the existence of an OBC eigenmode at the complex admittance value given by the $\mathrm{Re}\mathcal{J}$ and $\mathrm{Im}\mathcal{J}$ coordinates of the column. For each eigenmode, the $z$ coordinate of the dots denote the spatial position on the chain, and the color of the dot at each location the relative node voltage magnitude at the node with darker colors indicating larger magnitudes. at the base of each plot show the OBC admittance spectra OBC and PBC admittance spectra. c. The admittance spectra collapse and become identical under OBC and PBC at $C_{\mathrm{n}1}=C_{\mathrm{n}1}^{\mathrm{critical}}$. The NHSE vanishes at $C_{\mathrm{n}1}=C_{\mathrm{n}1}^{\mathrm{critical}}$. However, the two topological edge-states are still localized at the boundaries. d. The OBC and PBC exhibit very different spectra at $C_{\mathrm{n}1}>C_{\mathrm{n}1}^{\mathrm{critical}}$, at which all voltage eigenstates localization has switched to the right-most node. The same values of $C_2$, and $C_{\mathrm{n}2}$ as panel a are used in panels b to d. $C_1$ was fixed to 0.7 $\mu\mathrm{F}$, and $C_{\mathrm{n}1}$ was set to $0.55C_{\mathrm{n}1}^{\mathrm{critical}}$,  $C_{\mathrm{n}1}^{\mathrm{critical}}$, and $1.8C_{\mathrm{n}1}^{\mathrm{critical}}$ in b, c, and d, respectively. 
 }
\label{gFig2}
\end{figure*}

To better illustrate the aforementioned predictions for the localization of the skin modes in multiple non-reciprocal segments, we first turn to the generalized non-Hermitian SSH chain circuit shown in Fig. \ref{gFig1}a, with $N=2$ nodes per unit cell. The  two nodes are labelled $A$ and $B$ (the unit cell is marked in the dashed rectangle in Fig. \ref{gFig1}), where both the intra-cell and inter-cell couplings linking them are unequal in the backward and forward directions. Explicitly, we consider a stronger coupling in the backwards direction compared to the forward coupling in the inter-cell hopping (i.e. for positive $C_1$ and $C_{\mathrm{n}1}$, $C_1-C_{\mathrm{n}1} < C_1+C_{\mathrm{n}1}$) while the intra-cell couplings are weaker (stronger) in backward (forward) direction (i.e. for positive $C_2$ and $C_{\mathrm{n}2}$, $C_2+C_{\mathrm{n}2} > C_2-C_{\mathrm{n}2}$). The usual Hermitian SSH circuit can be recovered simply by setting $C_{\mathrm{n}1}=C_{\mathrm{n}2}=0$, which ensures that the forward and backward couplings are the same throughout the circuit lattice. In a physical circuit, which we simulate via LTspice, the asymmetric coupling in the $i$th segment is realized by connecting a capacitor of capacitance $C_i$ in parallel to a unity-gain operational amplifier (LT1056) and a capacitor that breaks the Hermiticity of the system and has capacitance $C_{\mathrm{n}i}$, which is half of the imbalance between the forward and backward couplings. Note that the sign of the non-Hermitian part of the couplings $C_{ni}$ can be flipped by reversing the biasings of the operational amplifier in the inter-cell and intracell segment couplings with respect to each other. This sign flipping of $C_{\mathrm{n}i}$ in the asymmetric coupling  is also known as negative impedance converter with current inversion (INIC) \cite{imhof2018topolectrical,rafi2021non}. For notational convenience, we  denote $(i \omega)^{-1}L(\beta, \omega)$ as $H(\beta,\omega)$ and call it the ``normalized Laplacian'' where $\omega$ is the frequency of the driving AC signal:
\begin{widetext}
\begin{equation}
 H(\beta,\omega)  = 
\begin{pmatrix}
 (\omega^2 L-C_2) & (C_1 -C_{\mathrm{n}1})+(C_2 -C_{\mathrm{n}2}) \beta\\
(C_1 +C_{\mathrm{n}1})+(C_2 + C_{\mathrm{n}2}) \beta^{-1} & (\omega^2 L-C_2)
\end{pmatrix},
\label{eq1}
\end{equation}
\end{widetext}
$H(\beta,\omega)$ is reminiscent of the Hamiltonian in quantum mechanics although strictly, it cannot be interpreted as a Hamiltonian because it is not a time evolution operator.)  The diagonal terms of Eq. \ref{eq1} represent a net shift in the eigenvalues of the Laplacian and can be adjusted by varying the frequency. At the resonant frequency of $\omega_r=\frac{1}{\sqrt{L_{\mathrm{g}} C_2} }$, the  diagonal terms vanish. The OBC admittance band dispersion as a function of $C_1$ at the resonant frequency is shown in Fig. \ref{gFig2}a. Zero-admittance edge states emerge within the parameter range of $-\sqrt{C_2^2-C_{\mathrm{n}2}^2+C_{\mathrm{n}1}^2} < C_1<\sqrt{C_2^2-C_{\mathrm{n}2}^2+C_{\mathrm{n}1}^2}$. Additionally, the solutions of $\beta$ in $|H(\beta, \omega) - \mathcal{J}\mathbf{I}_2|= 0$ determine the localization of the skin eigenmodes of the open chain, where $\mathbf{I}_2 $ is the $2 \times 2$ identity matrix. Explicitly, the bulk voltage modes of the circuit chain under OBC are localized at one of the boundary nodes with $|V_x| \approx e^{-\kappa x}$, where
\begin{equation}
\kappa = -\frac{1}{2} \ln \left| \frac{ (C_1-C_{\mathrm{n}1})(C_2+C_{\mathrm{n}2})}{(C_1+C_{\mathrm{n}1})(C_2-C_{\mathrm{n}2}) }  \right|.
\label{eq4}
\end{equation} 
Any non-zero value of $\kappa$ will result in an extensive voltage eigenstates localization. However, $\kappa$ goes to zero if $C_{\mathrm{n}1}$ reaches its critical value of 
\begin{equation}
C_{\mathrm{n}1}=C_{2\times 2}^{\mathrm{critical}}=\frac{C_1 C_{\mathrm{n}2}}{C_2} \label{crit2}
\end{equation}
(see Methods section for detailed derivation), at which the GBZ becomes the usual Bloch BZ of a unit circle on the complex plane, and the extensive localization of the voltage eigenstates vanishes. In contrast, the inverse localization length $\kappa$ is positive if $C_{\mathrm{n}1}> C_{2\times 2}^{\mathrm{critical}}$ and negative when $C_{\mathrm{n}1}< C_{2\times 2}^{\mathrm{critical}}$ for an open TE chain. As a result, all voltage eigenstates are localized at the left most node if $C_{\mathrm{n}1}< C_{2\times 2}^{\mathrm{critical}}$ (see Fig. \ref{gFig2}b) and  at the right most node if $C_{\mathrm{n}1}> C_{2\times 2}^{\mathrm{critical}}$ (see Fig. \ref{gFig2}d). As expected, when $C_{\mathrm{n}1}\neq C_{2\times 2}^{\mathrm{critical}}$, the OBC and PBC spectra for the admittance eigenvalues are very different. However, the PBC and OBC admittance dispersions become real and identical when $C_{\mathrm{n}1}= C_{2\times 2}^{\mathrm{critical}}$.   Therefore, the usual BBC is restored at the critical value of the non-Hermitian parameter $C_{\mathrm{n}1}$ without the accumulation of skin modes. However, the topological edge modes are still localized near the boundary nodes at the critical condition (see Fig. \ref{gFig2}c).
We next extend our analysis by considering a circuit consisting of a  4-node unit cell with asymmetric couplings between its 4 segments, as shown in Fig. \ref{gFig1}c. (Note that we have now made the biasing of the op-amps in the circuit uniform to simplify the resulting expressions. A negative value of $C_{\mathrm{n}i}$ indicates that the biasing of the $i$th op-amp is reversed.  ) 
For this four-segment system,
\begin{equation}
\beta_{4 \times 4}= \sqrt{\prod_i^4 \frac{ |C’_i + C’_{\mathrm{n}i}|}{|C’_i - C’_{\mathrm{n}i}|}} ,
\label{eq5}
\end{equation}  
(see Methods) which follows the same general form as Eq. \ref{eqb}. The new value for, say $C’_{\mathrm{n}4}$ at which both PBC and OBC spectra overlap and have the identical eigenvalues distributions occurring at
\begin{equation}
C’_{\mathrm{n}4} = -\frac{C'_4( C'_1( C'_3C'_{\mathrm{n}2} + C'_2C'_{\mathrm{n}3}) + C'_{\mathrm{n}1}(C'_2C'_3 + C'_{\mathrm{n}2}C'_{\mathrm{n}3}) )}{C'_3( C'_1C'_2 + C'_{\mathrm{n}1}C'_{\mathrm{n}2}) + C'_{\mathrm{n}3}(C'_1C'_{\mathrm{n}2} + C'_2 C'_{\mathrm{n}1})}.
\label{eqcritical}
\end{equation}
Furthermore, similar to the two-segment case, the conventional BBC is restored with no skin mode accumulation except for any topological edge modes that may exist depending on the circuit parameters. 
\subsection{Square-root topological lattice in TE circuit with multiple asymmetric couplings branches} 
As mentioned earlier, the circuit in Fig. \ref{gFig1}a can be made (by appropriate setting of the capacitor values) to harbour topological SSH-like edge states in addition to the skin modes \cite{rafi2021topological}. Interestingly, with proper tuning of the coupling capacitors, the model circuit with 4-node unit cells depicted in Fig. \ref{gFig1}c can be transformed to a square-root topological lattice \cite{arkinstall2017topological,kremer2020square,song2020realization,ezawa2020systematic} whose parental topological insulator corresponds to the 2-node circuit of Fig. \ref{gFig1}a. The square-root lattice can be made to inherit not only the topological properties, but also the NHSE characteristics from its parental topological insulator in Fig. \ref{gFig1}a  using the approach in Ref. \onlinecite{ezawa2020systematic}. The hallmark of a square-root topological lattice is the emergence of the boundary modes at non-zero admittance (equivalent to the non-zero energy in the condensed matter). Although such non-zero admittance boundary states may seem accidental, their origin are actually topological in nature \cite{arkinstall2017topological}. The well-known zero-admittance topological boundary states  can be retrieved simply by squaring the Laplacian of a square-root topological lattice.  In brief, given a $N \times N$ Hamiltonian $H_{\mathrm{original}}$ containing $M$ hoppings between the lattice sites, a normalized $(N+M)\times(N+M)$ Hamiltonian $H_{\mathrm{root}}$ can be constructed by introducing a new node between every pair of nodes that are coupled together in $H_{\mathrm{original}}$. The coupling between the new node and the original pair of nodes  is the square root of the coupling between the original pair of nodes. The $H_{\mathrm{root}}$ thus constructed is the square root of $H_{\mathrm{original}}$ in the sense that $H_{\mathrm{root}}^2 =  (H_{\mathrm{original}} + c \mathbf{I}_{N }) \oplus H_{\mathrm{residual}}$ where the normalized residual Hamiltonian $H_{\mathrm{residual}}$ is a $M \times M$ Hamiltonian containing only couplings between the newly introduced nodes, $c$ is a non-zero constant calculated from the coupling magnitudes in $H_{\mathrm{root}}$, and $\mathbf{I}_{N}$ is the $N \times N$ identity matrix. The $c$ term leads to the emergence of \textit{non-zero} energy topological states in $H_{\mathrm{root}}$ that correspond to the zero-energy states in $H_{\mathrm{original}}$.  
We realize the Laplacian analogous to the $H_{\mathrm{root}}$ of the circuit in Fig. \ref{gFig1}a  by setting the following capacitances in the circuit in Fig. \ref{gFig1}c to the following values:
\begin{eqnarray}
&&C’_1 = C’_2 = \frac{1}{2}(\sqrt{C_1+C_{\mathrm{n}1}}+\sqrt{C_1-C_{\mathrm{n}1}}), \\
&& C’_{\mathrm{n}1} = C’_{\mathrm{n}2}  = \frac{1}{2}(\sqrt{C_1-C_{\mathrm{n}1}}-\sqrt{C_1+C_{\mathrm{n}1}}), \\ 
&&C’_3 = C’_4 =\frac{1}{2}(\sqrt{C_2+C_{\mathrm{n}2}}+\sqrt{C_2-C_{\mathrm{n}2}}), \\
&&C’_{\mathrm{n}3} = C’_{\mathrm{n}4} = \frac{1}{2}(\sqrt{C_2+C_{\mathrm{n}2}}-\sqrt{C_2-C_{\mathrm{n}2}}).
\label{eq6}
\end{eqnarray} 
In Fig. \ref{gFig1}c, ($A_0$, $B_0$) correspond to the original nodes ($A$, $B$) in Fig. \ref{gFig1}a, while  $A_i$ and  $B_i$ are respectively the nodes added astride the intra- and inter-cell coupling between the $A$ and $B$ nodes. The resultant normalized Laplacian ($H_{\mathrm{root}}$) of the modified circuit in Fig. \ref{gFig1}c satisfies the following  at resonant frequency (see the Methods section for the detailed derivation):
\begin{align}
&H_{\mathrm{root}}^2 = \nonumber \\  
&\left(\left(\sqrt{C_1^2-C_{\mathrm{n}1}^2}+\sqrt{C_2^2-C_{\mathrm{n}2}^2} \right) \mathbf{I}_{2}+ H_{\mathrm{resonant}} (\beta, \omega)\right)  \nonumber \\
&\oplus H_{\mathrm{residual} } (\beta),
\label{eq8}
\end{align}
where $H_{\mathrm{resonant}} (\beta, \omega)$ denotes the normalized circuit Laplacian of Eq. \ref{eq1} at resonant frequency and $H_{\mathrm{residual}} (\beta)$ is a $2 \times 2$ matrix that takes the form of 
\begin{widetext}
\begin{align}
&H_{\mathrm{residual}} (\beta)= \nonumber \\
&\begin{pmatrix}
2 \sqrt{C_1^2-C_{\mathrm{n}1}^2} & \sqrt{(C_1 -C_{\mathrm{n}1})(C_2 +C_{\mathrm{n}2})}+\sqrt{(C_1 +C_{\mathrm{n}1})(C_2 - C_{\mathrm{n}2})} \beta\\
\sqrt{(C_1 +C_{\mathrm{n}1})(C_2 - C_{\mathrm{n}2})}+\sqrt{(C_1 - C_{\mathrm{n}1})(C_2 + C_{\mathrm{n}2})} \beta^{-1} & 2 \sqrt{C_2^2-C_{\mathrm{n}2}^2}
\end{pmatrix},	
\label{eq9} 
\end{align}
\end{widetext}
\begin{figure*}[ht!]
\centering
\includegraphics[width=0.9\textwidth]{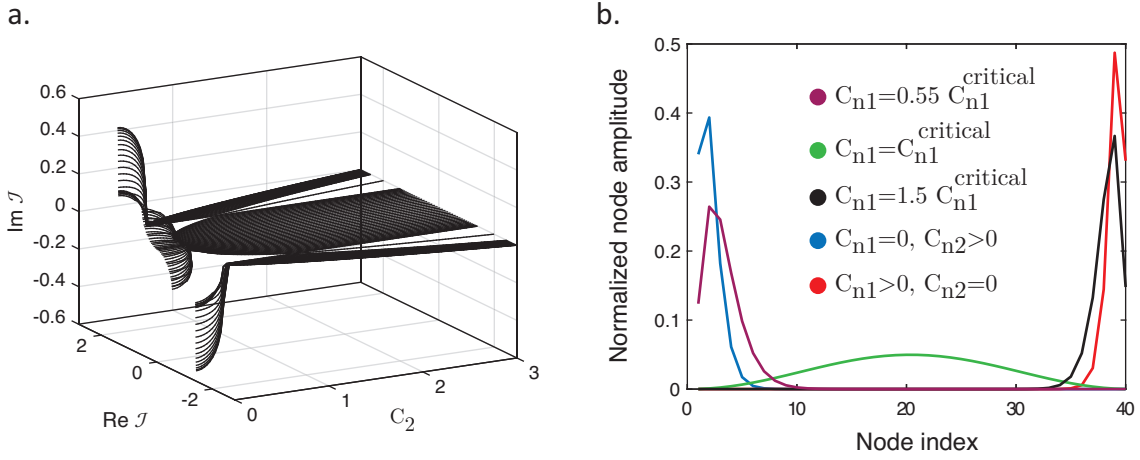}
\caption{Admittance dispersion and node voltage localization in a square-root topological circuit with multiple asymmetric couplings at resonant frequency.  a. Admittance spectrum as a function of $C_2$ under OBC for the circuit Laplacian (see Eq. \ref{eq8}). 
Parameters used: $C_1=1$ $\mu\mathrm{F}$,  $C_{\mathrm{n}1}=0.2$ $\mu\mathrm{F}$, and $C_{\mathrm{n}2}=0.3$ $\mu\mathrm{F}$. Note that the topological edge-states occur at a non-zero admittance of  $\mathcal{J}=\pm \sqrt{\sqrt{C_1^2-C_{\mathrm{n}1}^2}+ \sqrt{C_2^2-C_{\mathrm{n}2}^2}}$. 
c. Node localization of arbitrarily chosen modes under single nonzero (i.e., only one of $C_{\mathrm{n}1}$ or $C_{\mathrm{n}2}$ is non-zero)and multiple asymmetric (i.e., $C_{\mathrm{n}1}$ and $C_{\mathrm{n}2}$ are non-zero) coupling. When asymmetric coupling is present in only one segment, the eigenmode localization is changed between the left or right edges simply by switching the sign of the non-Hermitian parameter $C_{\mathrm{n}i}$. However, when asymmetric coupling is present in multiple segments, the localization direction is determined by the relative strength of $C_{\mathrm{n}1}$ to its critical value $C_{\mathrm{n}1}^{\mathrm{critical}}$. The coupling conditions for the different eigenstates are indicated by the lines of various color lines. Common parameters: $C_1=0.7$ $\mu\mathrm{F}$ and  $C_2=1$ $\mu\mathrm{F}$. ($C_{\mathrm{n}1}, C_{\mathrm{n}2}=(0, 0.7), (0.6, 0), (0.5 C_{\mathrm{n}1}^{\mathrm{critical}}, 0.7),(C_{\mathrm{n}1}^{\mathrm{critical}}, 0.7), (1.55 C_{\mathrm{n}1}^{\mathrm{critical}}, 0.7)$ $\mu\mathrm{F}$) for blue, red, magenta, green, and black curves, respectively. 
}
\label{gFig3}
\end{figure*}
A hallmark of a square-root TIs is the emergence of the mid-gap edge-states at non-zero admittance (see Fig. \ref{gFig3}a, \ref{gFig3}b). More precisely, the zero-admittance topological states in the original model in Fig. \ref{gFig1}a are shifted to a finite admittance of $\mathcal{J}=\pm \sqrt{\sqrt{C_1^2-C_{\mathrm{n}1}^2}+ \sqrt{C_2^2-C_{\mathrm{n}2}^2}}$ for $|C_1|> \sqrt{C_2^2-C_{\mathrm{n}2}^2 + C_1^2}$.
Finally, the effect of various asymmetric couplings on the skin mode localization is summarized in Fig. \ref{gFig3}c. When asymmetric coupling is present only along one segment (i.e., $C_{\mathrm{n}1}\neq 0, C_{\mathrm{n}2}= 0$, or $C_{\mathrm{n}2}\neq 0, C_{\mathrm{n}1}= 0$),  the eigenmodes are localized to the left or right boundaries (see blue and red color lines) with the respective inverse decay lengths of $\alpha_{\mathrm{a}}= \sqrt{\frac{C_1-C_{\mathrm{n}1}}{C_1+C_{\mathrm{n}1}}}$ and $\alpha_{\mathrm{b}}= \sqrt{\frac{C_2+C_{\mathrm{n}2}}{C_2-C_{\mathrm{n}2}}}$. However, when both asymmetric couplings are present, the overall skin modes are accumulated with the inverse decay length of $\alpha_{\mathrm{a}}  \alpha_{\mathrm{b}}$ (see black and magenta color lines), and the NHSE vanishes when $\alpha_{\mathrm{a}}$ and $\alpha_{\mathrm{b}}$ are inverses of each other (see green color line).
 \subsection{Experimental proposal}
A test of whether our theoretical TE models and circuits can be realized via physical circuits is to simulate them using LTSpice simulations based on realistic circuit components. In this section, we now provide the details for the experimental verification of the TE circuits. To evaluate the complex admittance spectra, we supply an external alternating current $I_i$ of a fixed magnitude (say, 1 mA) at the angular frequency $\omega$ at a particular node $i$ and measure the resulting voltages at all nodes in the circuit chain using a lock-in amplifier. The voltage at node $j$, which we denote as $V_{j}^i$, is related to the current supplied to node $i$ by 
 \begin{equation}
 V_{j}^i/I_{i}= L^{-1}_{ij},
 \label{eqexper}
\end{equation}  
where $i$ and $j$ denote the node indices and run from $1$ to $(n \times m)$ with $n$ and $m$ represent the number of nodes per unit cell and the number of unit cells in the chain, respectively. Eq. \eqref{eqexper} implies that all the matrix elements of $L^{-1}_{ij}$ can be obtained by iterating $i$ through all the nodes and measuring the resulting voltages in all the nodes for each value of $i$. After obtaining all the matrix elements of $L^{-1}$, the inverse of the $L^{-1}$ can be taken to obtain the admittance matrix for a given set of parameters. To obtain the PBC spectra, the nodes at the two ends of the finite-length chain are coupled together with the same strength as the inter-unit cell coupling, and  the above procedure is repeated for obtaining matrix elements of $L$. One way to experimentally obtain the non-Hermitian skin effect (NHSE) node voltage profile for a given eigenmode is to ground all the nodes of the circuit with a capacitance equal to the corresponding eigenvalue of that eigenmode and connect a voltage supply at an arbitrary node which has a non-zero amplitude in the eigenmode. The resulting voltage profile in the remaining nodes will then be proportional to the voltage profile of the normalized eigenmode. 
At the same time, it should be noted in the realization of asymmetric coupling via an INIC, precautions should be taken to avoid the possible breakdown of the circuit due to highly non-linear behaviours of the op amps when the threshold values for linearity are exceeded. To avoid such unwanted behaviour of the op amps, it has to be ensured that the non-Hermitian system contains certain loss to avoid instabilities and divergence of the INICs (see Supplementary section) for details).
\begin{figure*}[ht!]
\centering
\includegraphics[width=0.8\textwidth]{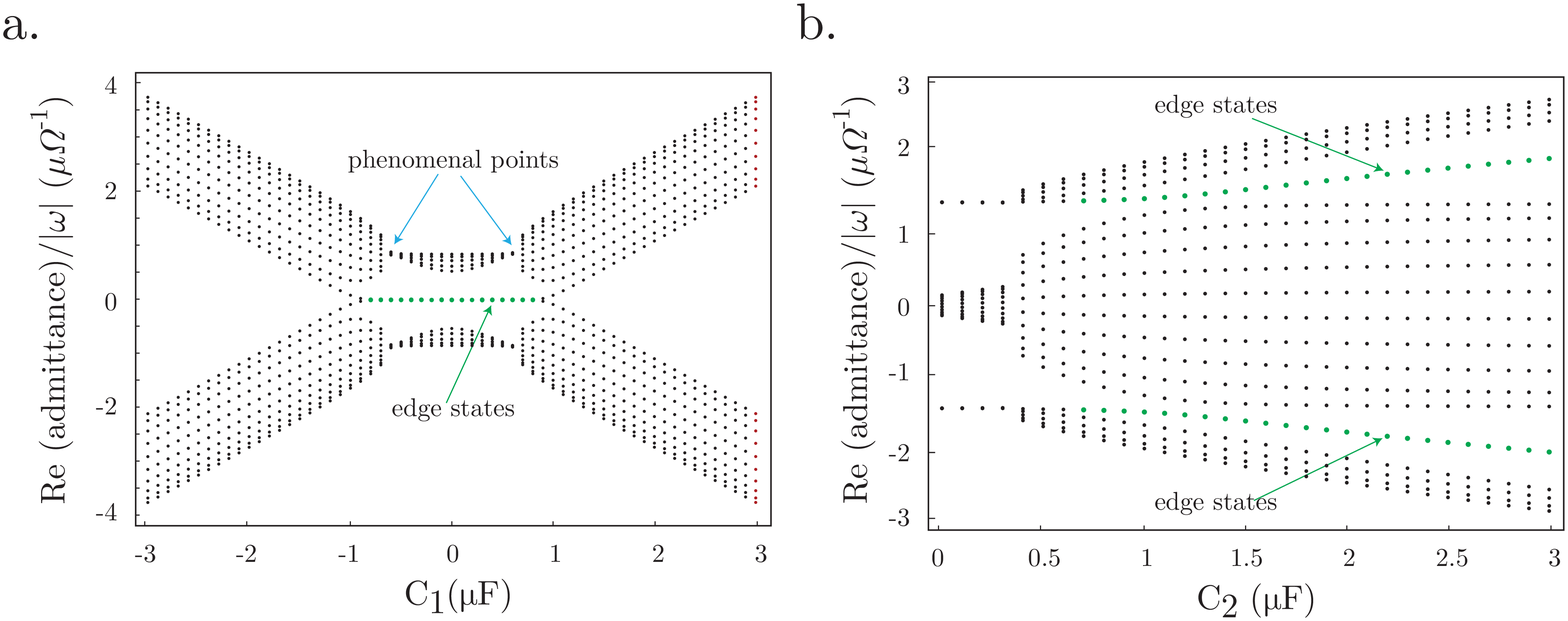}
\caption{Admittance band dispersions obtained via LTspice simulations of the circuit lattice with multiple asymmetric couplings. a. Real part of the resonant admittance dispersion obtained from LTspice simulations for the circuit described by Eq. \ref{eq1} under OBC conditions with 10 unit-cells. The edge states are characterized by their flattened zero-admittance dispersion. Circuit parameters used: $C_2=1$ $\mu\mathrm{F}$,  $C_{\mathrm{n}1}=0.6$ $\mu\mathrm{F}$, $C_{\mathrm{n}2}=0.5$ $\mu\mathrm{F}$, $L=100$ $ \mu\mathrm{H}$, $C_a =1 $ $\mu\mathrm{F} $ and $f_r = 15.9155$ kHz. b. Real part of admittance spectra under OBC for a square-root circuit described by Eq. \ref{eq8} with 5 unit-cells at resonant frequency. Clearly, non-zero admittance edge states appear in the bulk admittance gap. The coupling asymmetries are implemented by using operational amplifiers with the LTspice model number LT1056. Circuit parameters used: $C_1=1$ $\mu \mathrm{F}$,  $C_{\mathrm{n}1}=0.2$ $\mu \mathrm{F}$, $C_{\mathrm{n}2}=0.3$ $\mu \mathrm{F}$, $L=100 $ $\mu \mathrm{H}$, $C_a =1 $ $\mu \mathrm{F} $ and $f_r = 5.264$ kHz.  }
\label{gFig4}
\end{figure*} 
\begin{figure*}[ht!]
\centering
\includegraphics[width=0.9\textwidth]{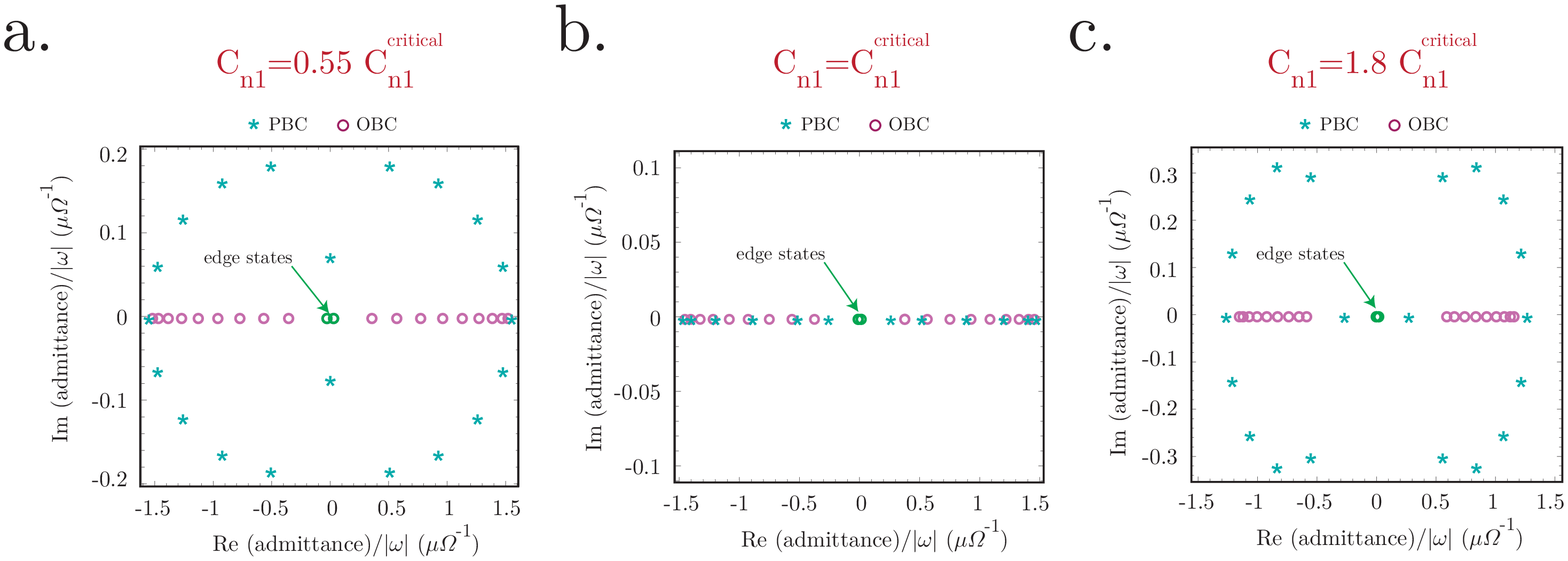}
\caption{Simulated PBC-OBC admittance spectra of the circuit described by Eq. \ref{eq1} at different levels of non-Hermiticity. a. Resonant admittance dispersion obtained from LTspice simulations when $C_{n1}$ is set to less than of the critical non-Hermiticity of $C_{\mathrm{n}1}^{\mathrm{critical}}= C_2^{-1} C_1 C_{\mathrm{n}2}$ (i.e., $C_{n1}=0.55 C_{\mathrm{n}1}^{\mathrm{critical}}$). b. The OBC and PBC admittance spectra in a critically tuned circuit lattice , i.e., one in which $C_{n1}= C_2^{-1} C_1 C_{\mathrm{n}2}$. The OBC and PBC spectra collapse and become coincident to each other, indicating the annihilation of the skin modes. c. The OBC and PBC admittance spectra when $C_{n1}$ exceeds $C_{\mathrm{n}1}^{\mathrm{critical}}$ (i.e., $C_{n1}=1.8 C_{\mathrm{n}1}^{\mathrm{critical}}$). The star and dot curves illustrate the evolution of the PBC and OBC spectra, respectively. Circuit parameters used: $C_1=0.7$ $\mu \mathrm{F}$, $C_2=1$ $\mu \mathrm{F}$,  $C_{\mathrm{n}1}=0.6$ $\mu \mathrm{F}$, $C_{\mathrm{n}2}=0.5$ $\mu \mathrm{F}$, $L=100$ $ \mu \mathrm{H}$, $C_a =1 $ $\mu \mathrm{F} $ and $f_r = 15.9155$ kHz. 
}
\label{gFig5}
\end{figure*} 
To verify the predictions in the previous sections, we performed LTspice simulation on circuit configurations with multiple asymmetric couplings, such as those depicted in Fig. \ref{gFig1}, and obtained their admittance spectra using the above procedure. The obtained admittance spectra from the LTspice simulations are consistent with the theoretical results based on the Laplacian formalism for both the generalized SSH circuit (cf. Figs. \ref{gFig4}a and \ref{gFig2}a)  and the square-root circuit lattice (cf. Figs. \ref{gFig4}b and \ref{gFig3}a)  when the frequency of the AC current signals in the LTspice simulation were set at the resonant frequencies of the respective circuits. Fig. \ref{gFig5} shows the LTspice simulation results analogous circuits to those in Fig. \ref{gFig2} with different multiplicative factors between the non-Hermitian parameter $C_{\mathrm{n}1}$ and its critical value. We obtained distinct admittance eigenvalue spectra under OBC and PBC conditions at resonant frequency when $C_{n1}$  deviates from $C_{\mathrm{n}1}^{\mathrm{critical}}$. This confirms the presence of skin modes (see Fig. \ref{gFig5}a, c). Furthermore, at $C_{n1} = C_{\mathrm{n}1}^{\mathrm{critical}}$, both the simulated PBC and OBC admittance eigenvalue spectra become coincident with each other (see Fig. \ref{gFig5}b) as predicted in Section \ref{sec:sqrtlattice} . Finally, we can also reproduce the eigenskin-modes distribution via LTspice simulations (see supplementary section ). The excellent matching of the realistic LTSpice simulation results with our theoretical predictions strongly suggests the experimental validity of our studied models.     
\section{Conclusion}
We analyzed the influence of the interplay between asymmetric coupling configurations in a non-Hermitian TE circuit on the onset and profile of the NHSE, i.e., extreme localization of the eigenmodes, both analytically as well as via realistic LTSpice simulations. Mathematically, the NHSE occurs in a TE circuit when the electrical couplings are non-reciprocal. We derived the general expression for NHSE in a TE system with multiple asymmetric couplings per unit cell. Our analysis reveals the existence of a critical value for a given non-Hermitian circuit parameter, at which that the NHSE is cancelled out, such that the node voltages are no longer localized at the boundaries even in the presence of non-reciprocity. Furthermore, our general model can be extended to a square-root TI with the appropriate choice of the coupling parameters. Such a square-root TI with multiple asymmetric couplings exhibits the unconventional characteristic of having topological states which are located at non-zero values of energy whilst retaining the edge localization. Finally, we proposed a scheme for the experimental implementation of TE version of our models which is confirmed by LTSpice simulation. 
Because our lattice models are described by only nearest-neighbour couplings, these models can also be realized in other platforms besides TE circuits such as optical \cite{longhi2020stochastic}, photonic \cite{zhu2020photonic,song2020two} and metamaterials \cite{schomerus2020nonreciprocal} systems. 
Our criterion for the overall NHSE strength, which takes the form of the geometric mean of the couplings, can be directly generalized to arbitrary long chains with any combination of coupling strengths. As such, it opens up a route towards the analysis of NHSE chains with inhomogeneity that extends across spatial regions, rather than being confined to single or a few couplings, thus  complementing our existing understanding on the role of spatial disorder in NHSE systems \cite{li2019geometric,li2020topological,li2021impurity} . \\
\section{Methods}
\subsection{GBZ solutions for the general model with four asymmetric couplings}
For the analytical derivation, we consider the general four-segment chain with four arbitrary asymmetric couplings segments per unit cell as shown in Fig. \ref{gFig1}c. The normalized Laplacian at resonant frequency can be written as 
\begin{align}
&H_{4 \times 4}(\beta)= \nonumber \\
&\begin{pmatrix}
0 & C_1 + C_{\mathrm{n}1} & 0 & (C_4 – C_{\mathrm{n}4})\beta \\
C_1 – C_{\mathrm{n}1} & 0 & C_2 + C_{\mathrm{n}2} & 0 \\ 
0 & C_2 – C_{\mathrm{n}2} & 0 & C_3+C_{\mathrm{n}3} \\
(C_4 + C_{\mathrm{n}4})\beta^{-1} & 0 & C_3 – C_{\mathrm{n}3} & 0 \end{pmatrix}.
\label{eq11}
\end{align}
The characteristic equation is then given by 
\begin{eqnarray} 
&&|(H_{4 \times 4 }(\beta)- \mathcal{J}\mathbf{I}_{4})|=0 \nonumber \\
  &\Rightarrow& \mathcal{J}^4 + \mathcal{J}^2 \Sigma + \Lambda = 0 \label{eq12}
,
\end{eqnarray}
where $\Sigma= \sum_{i}^4 (C_{\mathrm{n}i}^2 – C_i^2)$ and $\Lambda = [((C_2 + C_{\mathrm{n}2})(C_4 + C_{\mathrm{n}4}) – (C_1-C_{\mathrm{n}1})(C_3 – C_{\mathrm{n}3})\beta)(-((C_1 + C_{\mathrm{n}1}(C_3+C_{\mathrm{n}3})) + (C_2 – C_{\mathrm{n}2})(C_4 – C_{\mathrm{n}4})\beta)] \beta^{-1}$. To find the OBC solutions for $\mathcal{J}$, we solve Eq. \eqref{eq12} for $\beta$ and obtain
\begin{align}
\beta_{\pm}= &\frac{1}{2 \Gamma_{\mathrm{left}}}\Big( -(\mathcal{J}^4 +\mathcal{J}^2 \Sigma + \Delta ) \nonumber\\
&\pm \sqrt{(\mathcal{J}^4 +\mathcal{J}^2 \Sigma + \Delta )^2 -4 \Gamma_{\mathrm{left}} \Gamma_{\mathrm{right}}}\Big),
\label{eq13}
\end{align}
where, $\Gamma_{\mathrm{left}}= -\prod_{i=1}^4 (C_i – C_{\mathrm{n}i})$, $\Gamma_{\mathrm{right}}= -\prod_{i=1}^4 (C_i + C_{\mathrm{n}i})$ and $\Delta= (C_1^2 – C_{\mathrm{n}1}^2)(C_3^2 – C_{\mathrm{n}3}^2) + (C_2^2 – C_{\mathrm{n}2}^2)(C_4^2-C_{\mathrm{n}4}^2)$. The values of $\beta$ which fall on the GBZ are obtained by requiring that  $|\beta_{+}|=|\beta_{-}|$. This in turn requires that the discriminant of the characteristic equation to be an imaginary multiple of the coefficient of $\beta^2$ in  Eq. \ref{eq12}, i.e.,
\begin{equation}
(\mathcal{J}^4 +\mathcal{J}^2 \Sigma + \Delta )^2-4 \Gamma_{\mathrm{left}} \Gamma_{\mathrm{right}}= -\nu^2 (\mathcal{J}^4 +\mathcal{J}^2 \Sigma + \Delta )^2,
\label{eq14}
\end{equation} 
where $\nu \in \Re$. Therefore, the bulk admittance in the thermodynamic limit $ \mathcal{J}_\infty$ is given by
\begin{equation}
\mathcal{J}_\infty^2=\frac{1}{2}\left(-\Sigma + \sigma_1 \sqrt{-4 \Delta +\Sigma^2 + \sigma_2 \frac{8 \Gamma_{\mathrm{left}} \Gamma_{\mathrm{right}}}{(1+\nu^2)\Gamma_{\mathrm{left}} \Gamma_{\mathrm{right}}}}\right)
\label{eq15}
\end{equation}
where $\sigma_1$ and $\sigma_2$ independently take the values of $\pm 1$ (i.e., there are four independent combinations of $(\sigma_1,\sigma_2)$. )  
The bulk OBC admittance spectrum is then given by the locus of $\mathcal{J}_\infty$ as $\mu$ is varied from $-\infty$ to $+\infty$. 
Finally, the GBZ can be obtained by taking the geometric mean of the two distinct solutions of $\beta$ satisfying $|\beta_+|=|\beta_-|$, i.e., $\beta_{4\times 4} = |\sqrt{\beta_+\beta_-}|$ and the non-Bloch multiplication factor can be calculated as 
\begin{equation}
\alpha_{4\times 4} = \ln (\beta_{4\times 4})= \frac{1}{2} \sum_i^4 \ln \frac{|C_i-C_{\mathrm{n}i}|}{|C_i + C_{\mathrm{n}i|}}
\label{eq16}
\end{equation}
Now, the critical value of any one of the non-Hermitian parameters (i.e., the $C_{\mathrm{n}i}$s) as a function of the other non-Hermitian parameters and the coupling capacitors can be found by solving for the value of the desired parameter at $\alpha_{4\times 4}=1$. For example, the critical value of $C_{\mathrm{n}1}$ is explicitly given as 
 \begin{align}
 &C_{\mathrm{n}1}^{\mathrm{critical}}= \nonumber \\
& -\frac{ C_1 (C_3 (C_4 C_{\mathrm{n}2} + C_2 C_{\mathrm{n}4}) + C_{\mathrm{n}3}(C_2 C_4 + C_{\mathrm{n}2}C_{\mathrm{n}4})}{C_4 (C_2 C_3 + C_{\mathrm{n}2}C_{\mathrm{n}3}) + C_{\mathrm{n}4} (C_3 C_{\mathrm{n}2} + C_2 C_{\mathrm{n}3})}.
 \label{eq17}
 \end{align}
 Any deviation of $C_{\mathrm{n}1}$ from its critical value of $C^{\mathrm{critical}}_{\mathrm{n}1}$  will ensure the exponential localization of the voltage eigenstates and the emergence of NHSE as discussed in the Results section.
\subsection{Critical value for arbitrary number of segments} 
In the previous subsection, we explicitly derived the critical value for a system with four segments. Here, we generalize the derivation to a system with an arbitrary number $N$ of segments described by a Laplacian with the generic form of Eq. \eqref{LbetaC}.
We note that the only terms that contain $\beta$ in Eq. \eqref{LbetaC} are the inter-unit cell coupling terms at the end ends of the unit cell which contain $\beta$ and $1/\beta$ respectively. The characteristic equation for the system $|L(\beta) - \mathbf{I}_N| = 0$ is thus quadratic in $\beta$, and can be written as
\begin{eqnarray}
&& |L(\beta) - \mathcal{J}\mathbf{I}_N| = 0 \nonumber \\
&\Rightarrow& \left (\prod^N_i C^{\mathrm{right}}_i \right) + f(\mathcal{J})\beta + \left (\prod^N_i C^{\mathrm{left}}_i \right)\beta^2 = 0 \label{arbE1}
\end{eqnarray}
where $f(\mathcal{J})$ is a polynomial in $\mathcal{J}$ whose exact form is unimportant for the argument here. (The $f(\mathcal{J})$ for the four-segment system in the previous section is explicitly given by $f(\mathcal{J}) = \mathcal{J}^4 + \mathcal{J}^2\Sigma + \Delta$.) 
Dividing Eq. \eqref{arbE1} by $\prod^N_i C^{\mathrm{left}}_i$ gives 
\begin{equation}
	\beta^2 + \tilde{f}\beta + \prod^N_i \frac{C^{\mathrm{right}}_i}{C^{\mathrm{left}}_i} = 0 \label{arbE1a}
\end{equation}
where $\tilde{f} = (\prod^N_i (C^{\mathrm{left}}_i)^{-1}) f$. We note that on the GBZ, the two solutions for $\beta$ of the characteristic equation have the same magnitude $|\tilde{\beta}|$, which allows us to write the solutions as $\tilde{\beta}_\pm = |\tilde{\beta}|\exp(i\phi_\pm)$ for some phase factor $\exp(i\phi_\pm)$. Because the characteristic equation is a quadratic equation in $\beta$ which has the roots $|\beta|\exp(i\phi_\pm)$, the characteristic equation can be written as 
\begin{eqnarray}
	&& (\beta-|\tilde{\beta}|\exp(i\phi_+))(\beta-|\tilde{\beta}|\exp(i\phi_-)) =  0 \nonumber \\
	&\Rightarrow & \beta^2  - |\tilde{\beta}|(\exp(i\phi_+) + \exp(i\phi_-))\beta \nonumber \\
&& + |\tilde{\beta^2}|\exp(i(\phi_+ + \phi_-))  = 0 \label{arbE2}.  
\end{eqnarray}
Comparing the terms independent of $\beta$ on the left side of the equal sign in Eqs. \eqref{arbE1a} and \eqref{arbE2}, we read off that 
\begin{equation}
	|\tilde{\beta}|^2 = \left| \prod^N_i \frac{C^{\mathrm{right}}_i}{C^{\mathrm{left}}_i} \right|,
\end{equation} 
which is nothing but Eq. \eqref{eqb}.  We note that this OBC system is always equivalent to a Hermitian model with couplings $\sqrt{C_i^2-c_{ni}^2}$ via a nonunitary change of basis. This basis rotates each pair of asymmetric hoppings to their (Hermitian) geometric mean. That also explains why the OBC spectrum for the chain is always real. 
\subsection{ GBZ and $\mathcal{J}_\infty$ for the square-root topological lattice model with multiple asymmetric couplings}
 The bulk OBC admittance spectra and the GBZ for square root topological lattice with multiple asymmetric branches (see Fig. \ref{gFig1}c) can easily be deduced  via Eq. \ref{eq15} and Eq. \ref{eq13} respectively, after the parameters substitution described in Eq. \ref{eq6}.  The bulk OBC spectra under the thermodynamic limit can be expressed as 
\begin{widetext}
 \begin{equation}
 \mathcal{J}_{\mathrm{sqrt}}^2 = \sqrt{C_1^2 - C_{\mathrm{n}1}^2}+  \sqrt{C_2^2 - C_{\mathrm{n}2}^2} \pm \sqrt{C_1^2+ C_2^2-C_{\mathrm{n}1}^2-C_{\mathrm{n}2}^2 \pm 2\sqrt{\frac{(C_1^2 - C_{\mathrm{n}1}^2)(C_2^2 - C_{\mathrm{n}2}^2)}{1+ \nu^2}}}.
 \label{eq18}
\end{equation}   
 Similarly, the non-Bloch factor $\beta$ can be simplified for square root topological lattice as 
 \begin{eqnarray}
 \beta_{\mathrm{sqrt}}&=&\frac{1}{2(C_1 - C_{\mathrm{n}1})(C_2 - C_{\mathrm{n}2})}(2 \sqrt{\tau_1 \tau_2}-2(\tau_1+\tau_2)\mathcal{J}_{\mathrm{sqrt}}^2 +\mathcal{J}_{\mathrm{sqrt}}^4) \pm \nonumber \\ &&   \sqrt{2(\mathcal{J}_{\mathrm{sqrt}}^4)-\sqrt{\tau_1}\mathcal{J}_{\mathrm{sqrt}}^2)} \sqrt{4 \tau_1 +4 \sqrt{\tau_1 \tau_2}+ \mathcal{J}_{\mathrm{sqrt}}^2 (-4 \sqrt{\tau_1}-2 \sqrt{\tau_2}+ \mathcal{J}_{\mathrm{sqrt}}^2 ) },
	\label{eq20}
 \end{eqnarray}
\end{widetext} 
 with $\tau_1 = C_1^2 - C_{\mathrm{n}1}^2$ and  $\tau_2 = C_2^2 - C_{\mathrm{n}2}^2$. 
 
 \subsection{Characteristics of the square root topological lattice at off-resonant frequency with multiple asymmetric segments}\label{sec:sqrtlattice}
The normalized Laplacian of the square-root topological lattice at an arbitrary frequency after a reordering of the basis from ($A_0$, $A_{\mathrm{i}}$, $B_0$, $B_{\mathrm{i}}$) to ($A_0$, $B_0$, $A_{\mathrm{i}}$,$B_{\mathrm{i}}$) can be written as  
 \begin{equation}
 H_{\mathrm{sqrt}}( \beta, \omega_{off})= \begin{pmatrix}
(C_2-\omega^{-2}L_{\mathrm{g}}^{-1})\mathbf{I}_{2} & H_{2\times 2}^{\mathrm{Left}} (\beta)\\
H_{2\times 2}^{\mathrm{Right}}(\beta) & (C_2-\omega^{-2}L_{\mathrm{g}}^{-1})\mathbf{I}_{2}
\end{pmatrix},
 \label{eqsroff}
 \end{equation}
 with $H_{2\times 2}^{\mathrm{Left}} (\beta)=\begin{pmatrix}
\sqrt{C_1-C_{\mathrm{n}1}}& \sqrt{C_2-C_{\mathrm{n}2}}\beta\\
\sqrt{C_1+C_{\mathrm{n}1}} & \sqrt{C_2+C_{\mathrm{n}2}}
\end{pmatrix}$ and $H_{2\times 2}^{\mathrm{Right}} (\beta)=\begin{pmatrix}
\sqrt{C_1+C_{\mathrm{n}1}}& \sqrt{C_1-C_{\mathrm{n}1}}\\
\sqrt{C_2+C_{\mathrm{n}2}} \beta^{-1} & \sqrt{C_2-C_{\mathrm{n}2}}
\end{pmatrix}$. However, the relation in Eq. \ref{eq8} becomes invalid at off-resonant frequency. We instead have 
\begin{equation}
H_{\mathrm{sqrt}}^2( \beta, \omega_{off})= \begin{pmatrix}
H_{\mathrm{par}}^{\mathrm{off}}& H_{\mathrm{up}}^{\mathrm{off}}\\
H_{\mathrm{dn}}^{\mathrm{off}} & H_{\mathrm{res}}^{\mathrm{off}}
\end{pmatrix} ,
\label{sqeqb}
\end{equation}
where, $H_{\mathrm{par}}^{\mathrm{off}} = (\sqrt{C_1^2-C_{\mathrm{n}1}^2}+\sqrt{C_2^2-C_{\mathrm{n}2}^2}+(C_2-(\omega^2L_{\mathrm{g}})^{-1})^2 ) \mathbf{I}_{2}+ H_{\mathrm{resonant}} (\beta, \omega_{\mathrm{res}})$, $H_{\mathrm{residual}}^{\mathrm{off}}= (C_2-(\omega^{2}L_{\mathrm{g}})^{-1})^2 ) \mathcal{I}_{2\times 2} + H_{\mathrm{residual}} (\beta)$, $H_{\mathrm{up}}^{\mathrm{off}}=2 (C_2-(\omega^{2}L_{\mathrm{g}})^{-1}) H_{2\times 2}^{\mathrm{Left}} (\beta)$ and $H_{\mathrm{dn}}^{\mathrm{off}}=2 (C_2-(\omega^{2}L_{\mathrm{g}})^{-1}) H_{2\times 2}^{\mathrm{Right}} (\beta)$ respectively. At resonant frequency, Eq. \ref{sqeqb} reduces to the direct sum of the $H_{\mathrm{par}}$ and $H_{\mathrm{residual}}$, as shown in Fig. \ref{eq8}. Even at off-resonant frequency where $\omega \neq (C_2 L_{\mathrm{g}})^{-1/2}$, all four blocks of Eq. \ref{sqeqb} exhibit the same condition as Eq. \eqref{crit2} for  the NHSE. Therefore, the whole TE chain represented by Eq. \ref{sqeqb} will undergo the extreme voltage eigenstate localization with the same decay rate as the original non-reciprocal circuit array depicted in Fig. \ref{gFig1}a, and the NHSE behaviour remain the same. 
  
 \subsection{Breakdown of the NHSE at the critical points}
 The NHSE vanishes at the critical values of the non-Hermitian parameters when the admittance eigenvalue falls within a certain range. To show this, we consider the circuit in  Fig. \ref{gFig1}a as an example and solve the GBZs at the critical value of $C_{\mathrm{n}1}$ by  replacing $C_{\mathrm{n}1}= C_2^{-1} C_1 C_{\mathrm{n}2}$ in the solutions of $ \beta_{\pm,2\times2}$ obtained from the relation $ |\mathcal{J}_{2\times 2}- H (\beta, \omega_{\mathrm{res}})|=0$. We obtain
\begin{widetext}
 \begin{eqnarray}
 \beta_{\pm,2\times2}&=&\frac{1}{2C_1 (-C_2 + \frac{C_{\mathrm{n}1}^2}{C_2} )} \times ( C_1^2 +C_2^2 -\mathcal{J}_{2\times 2}^2 -C_{\mathrm{n}2}^2 - \frac{C_1^2 C_{\mathrm{n}2}^2}{C_2^2} \pm \nonumber \\ && \sqrt{-4C_1^2 (C_2-\frac{ C_{\mathrm{n}2}^2}{C_2})^2 + (-C_{2}^2 +\mathcal{J}_{2\times 2}^2 + C_{\mathrm{n}2}^2+ C_1^2 (-1+ \frac{ C_{\mathrm{n}2}^2}{C_2^2})) }     .
 \label{eq21}
 \end{eqnarray}
\end{widetext}
 The magnitude of the GBZs in Eq. \ref{eq21} will always be unity as long as $\mathcal{J}_{2\times 2}^2$ satisfies $\mathcal{J}_{\mathrm{cr}1}^2 < \mathcal{J}_{2\times 2}^2 < \mathcal{J}_{\mathrm{cr}2}^2$, where $\mathcal{J}_{\mathrm{cr}1}^2$ and $\mathcal{J}_{\mathrm{cr}2}^2$ represent the two extreme limit of the square of the admittance eigenenergy within which the GBZs fails completely. The two extreme limits of the $\mathcal{J}_{2\times 2}^2$ can  be evaluated by setting the discriminant of the characteristic polynomial of the Laplacian in Eq. \ref{eq1} to zero at the resonant frequency and setting $C_{\mathrm{n}1}$  at the critical value of $C_{\mathrm{n}1}= C_2^{-1} C_1 C_{\mathrm{n}2}$. Explicitly, we obtain
 \begin{equation}
 \mathcal{J}_{\mathrm{cr}1/\mathrm{cr}2}^2 = C_2^2-C_{\mathrm{n}2}^2 \mp 2 \sqrt{\frac{C_1^2 (-C_2^2+ C_{\mathrm{n}2}^2)}{C_2^2}} + C_1^2 (1- \frac{C_{\mathrm{n}2}^2}{C_2^2}),
 \label{eq22}
\end{equation} 
where, $\mp$ represents the lower and upper limits for $\mathcal{J}_{2\times 2}^2$ respectively.  Within these limits of $\mathcal{J}$, the resultant GBZ at the critical value of $C_{\mathrm{n}1}$ becomes the usual unit circle on the complex plane and the NHSE vanishes.  
\subsubsection*{Acknowledgements}
This work is supported by the Ministry of Education (MOE) Tier-II grant MOE2018-T2-2-117 (NUS Grant Nos. R-263-000-E45-112/R-398-000-092-112), MOE Tier-I FRC grant (NUS Grant No. R-263-000-D66-114), MOE Tier-I grant R-144-000-435-133, and other MOE grants (NUS Grant Nos. C-261-000-207-532, and C-261-000-777-532).

%

\end{document}